\documentclass[preprint]{aastex}
\usepackage{psfig}
\newcommand{\beq}{\begin{equation}}
\newcommand{\eeq}{\end{equation}}
\newcommand{\rem}[1]{{ }}

\bibpunct[,]{(}{)}{;}{a}{}{,}

\begin{document}

\title{Limits on thermal conduction in galaxy clusters}
\author{Mikhail V. Medvedev\altaffilmark{1,4}, 
Adrian L. Melott\altaffilmark{1,5}, 
Chris Miller\altaffilmark{2,6}, Donald Horner\altaffilmark{3,7}}
\altaffiltext{1}{Department of Physics and Astronomy, University of Kansas, 
Lawrence, KS 66045}
\altaffiltext{2}{Department of Physics and Astronomy, 
Carnegie Mellon University, Pittsburgh, PA 15213}
\altaffiltext{3}{Space Telescope Science Institute, Baltimore, MD 21218}
\altaffiltext{4}{medvedev@ku.edu}
\altaffiltext{5}{melott@kusmos.phsx.ukans.edu}
\altaffiltext{6}{chrism@cmu.edu}
\altaffiltext{7}{horner@stsci.edu}

\begin{abstract}
We have calculated lower limits for the global effective 
thermal conductivity in a sample of 165 Abell clusters.
We assumed that cluster X-ray luminosity is compensated by a
conductive heat flux which we then compare with 
an upper limit to the temperature gradient inferred from the 
cluster temperature and radius. This gives a lower limit to the 
thermal conductivity and therefore to the relative suppression 
from the Spitzer conductivity, $\kappa_{Sp}$. 
Not a single cluster requires super-Spitzer values of 
thermal conduction to balance the observed X-ray
luminosity. The suppression coefficient $f=\kappa/\kappa_{Sp}$ 
is clustered in a range $10^{-2} < f < 0.4$. A weak dependence
of $f$ versus $z$ is observed over $ 0 < z <0.41$.
Possible biases and/or selection effects are discussed.
\end{abstract}

\keywords{galaxies: clusters: general --- conduction}
\maketitle

\maketitle

\section{Introduction}

For some time it has been believed that the strong excess of X-ray
emission in the cores of many galaxy clusters represents a solid evidence of 
radiatively cooling gas which is in near hydrostatic equilibrium with a
surrounding medium.  As the cooling time of the core gas is rather
short, a gentle inward flow of gas should replenish the cooled gas presumably
condensing out.  \citet{Fabian-rev} contains a review of this picture.

This conventional paradigm of cooling flows turned
out to be grossly inconsistent with recent X-ray data on the 
central regions in galaxy clusters from {\it XMM-Newton} and 
{\it Chandra} \citep{Tamura+01,Fabian+01,Bohringer+01,Peterson+02}, which 
failed to find multi-temperature gas one expects in a cooling flow 
model. One of the possible explanations, --- extra heating via thermal
conduction flux --- was proposed by \citet{NM01}.
This mechanism requires sufficiently high values of the thermal 
conductivity, close to the Spitzer limit. Thus, magnetic fields 
permeating hot cluster gas may not suppress conductivity by 
a large factor. Theoretical considerations and numerical 
simulations of conduction in a turbulent gas agree that 
the suppression factor, $f=\kappa/\kappa_{Sp}$, is near unity, typically
around 1/3 \citep{NM01,MK01,Cho+03}. There is an indication that 
intermittency in turbulence (if present) may decrease $f$ by an additional 
factor of few to ten \citep{CM03}.

The idea of conduction-heated intra-cluster gas has been
tested using some available cluster data \citep{Voigt+02,%
Fabian+02,ZN03}. The inferred suppression factor was typically
in the range $1< f <1/10$, being consistent with the 
theoretical predictions. In the present paper we perform
similar analysis using a much larger sample of cluster data.

\section{The sample}

We use a sample of 165 Abell clusters \citep{abe58,aco89} 
with archived ASCA observations.
The clusters cover a wide range of richnesses ($0 \le {\rm R} \le 3$),
redshifts ($0.02 \le z \le 0.41$) and bolometric luminosities ($10^{43}$ to
$10^{46}$~ergs/s for $H_0 = 50$~km/s/Mpc).
The data were analyzed in semi-automated fashion and the cluster
temperatures were determined via spectral fits to the available ASCA data.  
We fit the spectra with a single temperature MEKAL plasma model using
XSPEC.  All four ASCA instruments (2 GIS + 2 SIS) were fit simultaneously
with the model parameters for each fit constrained to be the same,
except for the relative normalizations and the hydrogen column density.
The column density was fixed at the Galactic value for the GIS since it is
relatively insensitive below 1 keV but left as a free parameter for the
SIS since radiation damage to the CCDs manifests itself as spuriously high
column densities. For more information about the sample and the 
analysis, see \citet{Horner+03}.

The temperatures were determined for broad-band (0.8-10.0 keV)
fluxes measured within a specified extraction radius.  
The extraction radius was an estimate of the level at which the count rate
rises to about 5 sigma above the background level.  Note that this is the
GIS extraction radius.  The SIS extraction radius was chosen to be 0.72
times the GIS radius since the SIS has a smaller PSF and empirical tests
showed that this resulted in a similar signal-to-noise ratio for the SIS
data.
A comparison between our temperatures and  those obtained from independent 
measurements and presented  in the literature shows excellent agreement. 
The temperatures derived for such clusters are generally agree very well with
other isothermal fits, usually within about 10\%.
X-ray luminosities are bolometric, i.e., they are estimated from the best fit
model over the 0.01-50 keV range. The luminosities that we derive for the
ASCA catalog agree well with luminosities from ROSAT survey like the
BCS or NORAS with a scatter about the mean of about 25\%\footnote{
There is actually an offset with respect to older Einstein or EXOSAT
luminosities.  The ASCA luminosities are systematically higher by about
20\%.  This is probably due to some calibration issue.
}.

This cluster sample cannot be considered complete. The use of ASCA 
data puts some physical constraints on the types of clusters we have 
analyzed. For instance, the sensitivity of ASCA was such that no clusters
with T $<$ 5 keV are expected above $z \sim 0.4$. Likewise, the size of
the ASCA observing area prohibits the measurement of large (and usually high
temperature) clusters that are nearby. At best, this sample can be considered
a heterogeneous sampling of the Abell catalog. 

As with any cluster sample, the Abell cluster catalog suffers
from projection effects and incompletenesses. The purity of
are sample is not an issue for this analysis, since all of our
clusters have Xray and galaxy redshift information (to
ensure the reality of the systems). In terms of completeness,
we have decided to use only Abell clusters as their redshift
and richness distributions are well studied in the literature
\citep[ among others]{pew92,phg92,ebe96,jf99,mil+99,mil+02}.

We use the following cluster parameters available: the cluster redshift, $z$, 
the bolometric X-ray luminosity, $L_X$, the temperature with 90\% confidence 
limits, $T,\ T_{lo},\ T_{hi}$, and the extraction radius used for the 
spectrum fits, $R_{ext}$.

\section{The method}

To estimate thermal conduction we use the following method:
We assume that the cluster medium is static and is in thermal 
equilibrium, that is cooling of gas via Bremsstrahlung emission
is compensated by the conductive flux of heat from outer
parts of the cluster:
\beq
L(<r) = 4\pi r^2 \kappa\,({dT}/{dr}),
\label{L}
\eeq
where $L(<r)$ is the total luminosity within radius $r$.
We assume that $L\simeq L_X$. For this sample we have
no measurements of the temperature gradients.
Therefore we use $R_{ext}$ to estimate it. 
The gradient
scale may be larger or comparable to $R_{ext}$.
This implies an upper limit to the temperature gradient,
and accordingly a lower limit on the conductivity.
Indeed, the larger the conduction, the closer the cluster
to being isothermal and, hence, the smaller the 
temperature gradient. We use the following approximation 
to estimate the temperature gradient:
\beq
dT/dr\approx 0.4\, T/R_{ext}.
\label{dT}
\eeq
Here the factor 0.4 is an empirical factor obtained from 
clusters with high spatial resolution to ensure reasonable fits 
to the observed temperature gradients \citep{Fabian-rev,Fabian+02}.
Note that since the  numerical factor is smaller than unity, we obtain a 
more conservative lower limit on the effective $\kappa$. 
A similar technique has been used by \citet{Voigt+02} and 
\citet{Fabian+02} for smaller samples.
Note that an upper limit to the temperature gradient will imply a lower limit
on the thermal conductivity necessary to balance the energy losses due
to the X-ray emission. 

Finally, we calculate the suppression coefficient as
\beq
f=\kappa/\kappa_{Sp}
\eeq
obtained from eqs. (\ref{L}) and (\ref{dT}), and 
$\kappa_{Sp}=8.2\times10^{20} T_1^{5/2}$~erg/cm/s/keV
is the Spitzer thermal conductivity, and $T_1=T/(10~{\rm keV})$.
We estimated error-bars using $T_{lo}$ and $T_{hi}$
in place of $T$.

\section{Results}

\begin{figure}
\psfig{file=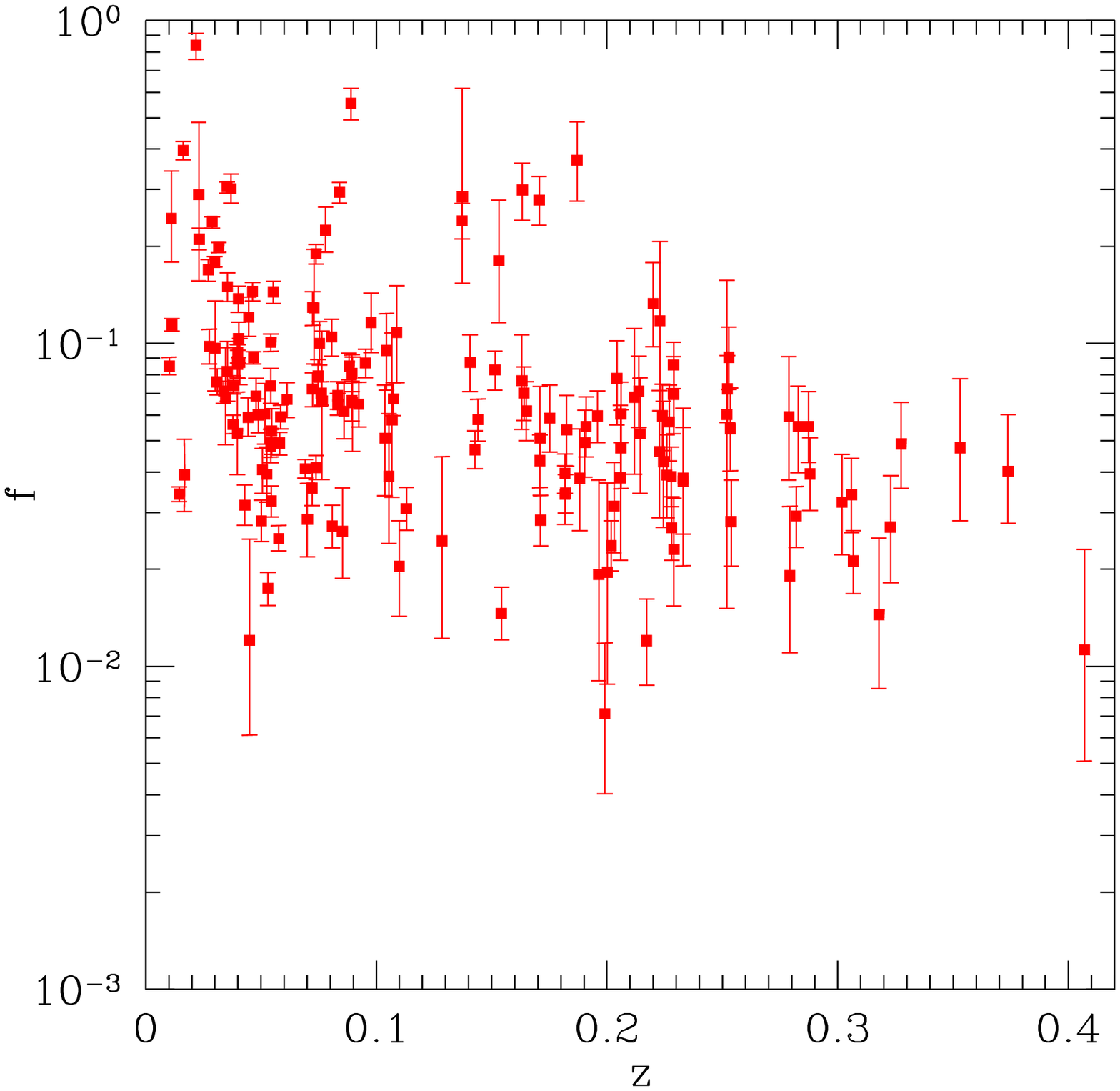,width=6in}
\caption{Suppression factors of thermal conduction in 
165 Abell clusters as a function of redshift.
Error bars result from 90\% confidence limits on $T$.  Errors in $z$ are
negligible for this purpose.
\label{f:kappa}}
\end{figure}

The result is shown in Fig. \ref{f:kappa}.
Suppression coefficients greater than unity would mean
that thermal conduction is insufficient to balance radiative
cooling; this sets a global constraint: $f<1$. In a {\em static} 
gas (i.e., without strong turbulent motions) with tangled magnetic 
fields, the constraint is more stringent, $f\le1/3$ \citep{NM01}.
Apparently, not a single cluster violates the first one and only three
violate the second one at the statistically significant level ($>90$\%).
The values of $f$ are clustered in the range
$10^{-2} < f < 0.4$.

We examined the distribution of $f$ against $z$.
We found a mild redshift dependence of $f$ over the range $ 0 < z <0.41$,
the slope is $-0.17\pm0.03$.
This result may however be affected by several biases and 
selection effects. More distant clusters in the sample (i) tend to
have larger $R_{ext}$, (ii) are more luminous, (iii) 
hence, have higher gas temperatures. 
The lack of strong $z$-dependence in our inferred values of $f$ may be
interpreted as additional support for the hypothesis that cool
cores are heated by thermal conductivity.

\section{Discussion}

Our lower limits on thermal conduction suppression are of course
dependent on the assumption that the X-ray emission in the core is
approximately balanced by heat flow from the surrounding medium.
Of course, other explanations are possible, such as localized heating by
compact objects in the cluster core.  Therefore our results should be
interpreted as a feasibility study for the thermal conductivity scenario.

Recently, numerical simulations with cooling \citep{Mo03} have been
presented which suggest ``cool cores'' form in a complex, time-dependent
merger process and do not require a steady flow.  This view is further
supported by \citet{Loken99}, who found that such objects
lie preferentially in clusters with a high density supercluster environment,
specifically having other clusters in high proximity.  Such an environment
would be expected to enhance the merger rate.  \citet{Mo03} comment
that their cluster temperature profiles are too steep (the cores too cool).
Since the cooling timescale is considerably shorter than the time between
major mergers, thermal conductivity should help resolve this discrepancy.

It has been believed (based on results from ordered magnetic fields) that
thermal conductivity is too weak to prevent runaway cooling and formation 
of bright, small-scale X-ray emitting regions in the
cores of galaxy clusters.  The proposal of \citet{NM01}, for enhanced
thermal conductivity in the presence of tangled magnetic fields is
consistent with the large data set studied here.  The data do not seem
to require heat transport by strong turbulent motion, reconnection
and turbulent resistivity, as proposed by \citet{CM03}.
The inclusion of thermal conductivity in numerical simulations of galaxy
cluster formation should be given a high priority in order to realistically
model the inner regions.

\section{Acknowledgments}

ALM wishes to acknowledge the support of the National Science Foundation
through grant AST0070702.

\end{document}